\documentclass{jaa}
\usepackage{natbib}
\usepackage{dcolumn}
\usepackage{graphicx}	
\usepackage{amsmath}	
\usepackage{amssymb}	
\usepackage{float}
\usepackage{natbib}
\usepackage{morefloats}
\usepackage{dcolumn}
\usepackage{amsmath}
\usepackage{latexsym}
\usepackage{longtable}
\usepackage{lipsum}
\usepackage{gensymb} 
\usepackage{adjustbox}
\usepackage{lscape}
\usepackage{empheq}
\usepackage{mathrsfs}
\usepackage{textcomp}
\usepackage{enumitem}   
\usepackage{gensymb}
\usepackage{hyperref}
\usepackage{graphicx}
\usepackage[caption=false]{subfig}
\usepackage{multirow}
\usepackage{longtable}
\usepackage{graphicx}

\begin{document}\sloppy

\title{Modeling the late time merger ejecta emission in short Gamma Ray Bursts}


\author{Ankur Ghosh\textsuperscript{1,2}, Kuntal Misra\textsuperscript{1}, C. S. Vaishnava \textsuperscript{3,4}, L. Resmi\textsuperscript{3}, K. G. Arun\textsuperscript{5}, Amitesh Omar\textsuperscript{1}, Dimple\textsuperscript{1} and N. K. Chakradhari\textsuperscript{2}}  

\affilOne{\textsuperscript{1}Aryabhatta Research Institute of observational sciencES (ARIES), Nainital, Uttarakhand, 263001, India.\\}
\affilTwo{\textsuperscript{2}Department of Physics, Pt. Ravi Shankar Shukla University, Raipur, Chattisgarh, 492010, India\\}
\affilThree{\textsuperscript{3}Indian Institute of Space Science and Technology (IIST), Thiruvanthapuram, Kerala, 695547, India.\\}
\affilFour{\textsuperscript{4}Physical Research Laboratory (PRL), Ahmedabad, Gujarat, 380009, India.\\}
\affilFive{\textsuperscript{5}Chennai Mathematical Institute, Siruseri, Tamilnadu, 603103, India.}


\twocolumn[{

\maketitle

\corres{ghosh.ankur1994@gmail.com}


\begin{abstract}
The short Gamma Ray Bursts (GRBs) are the aftermath of the merger of binary compact objects (neutron star -- neutron star or neutron star -- black hole systems). With the simultaneous detection of Gravitational Wave (GW) signal from GW 170817 and GRB 170817A, the much-hypothesized connection between GWs and short GRBs has been proved beyond doubt. The resultant product of the merger could be a millisecond magnetar or a black hole depending upon the binary masses and their equation of state. In the case of a magnetar central engine, fraction of the rotational energy deposited to the emerging ejecta produces late time synchrotron radio emission from the interaction with the ambient medium. In this paper, we present an analysis of a sample of short GRBs located at a redshift of $z \leq 0.16$ which were observed at the late time to search for the emission from merger ejecta.  Our sample consists of 7 short GRBs which have radio upper limits available from VLA and ATCA observations. We generate the model lightcurves using the standard magnetar model incorporating the relativistic correction. Using the model lightcurves and upper limits we constrain the number density of the ambient medium to be $10^{-5} - 10^{-3} cm^{-3}$ for rotational energy of the magnetar $E_{rot} \sim 5\times10^{51}$ erg. Variation of ejecta mass does not play a significant role in constraining the number density.

\end{abstract}

\keywords{gravitational waves -- surveys -- gamma-ray burst: general -- stars: magnetars -- stars: neutron.}
}]


\doinum{12.3456/s78910-011-012-3}
\artcitid{\#\#\#\#}
\volnum{000}
\year{0000}
\pgrange{1--}
\setcounter{page}{1}
\lp{1}

\section{Introduction}
The most accepted progenitor model for the origin of short GRBs is the  merger of binary compact objects (binary neutron stars (BNS) and neutron star - black hole (NS - BH)). These systems are also the prime candidates for producing gravitational waves and kilonovae. The joint detection of the GW 170817, GRB 170817A and AT2017gfo confirmed the connection between the three events \citep{2017ApJ...848L..13A, 2017ApJ...848L..24V, 2017ApJ...848L..15S, 2017PASA...34...69A, 2017Sci...358.1570D, 2017ApJ...848L..27T}.  
The most debated open question regarding the central engine of GRB is whether black hole production is necessary for the emergence of short GRB jet or the central engine could be a highly magnetized and rapidly spinning magnetar \citep{2001ApJ...552L..35Z, 2008MNRAS.385.1455M}. In case of a BNS merger, the nature of the remnant depends on the initial masses of the BNS and the equation of state of the NSs. The massive binaries ($\geq$ 3 $M_\odot$) will directly collapse to a black hole whereas the less massive BNS merger creates a transient state in between the merger and the production of black hole which is a millisecond magnetar. 
In the \textit{Swift} era the X-ray lightcurves from X-Ray Telescope (XRT, \citep{2004ApJ...611.1005G}) show a complex lightcurve morphology with an early time X-ray excess ($\leq$ 10 seconds), mid time flattening or plateau (10 - 1000 seconds) and late time X-ray excess followed by a sharp decay ($\geq$ 1000 seconds). The plateau phase in the X-ray lightcurves is thought to be powered by the magnetar \citep{2013MNRAS.430.1061R}.

The energy extraction from the central engine occurs via two channels. The first is through the emerging jets from the central engine which carries an enormous amount of energy and gets decelerated by the interaction with the nearby ambient medium. This emission is responsible for the prompt emission and afterglow of the GRBs. Another mode of energy injection is governed by the the isotropic ejecta emerging after the merger. As the ejecta is neutron rich, the matter released during the merger undergoes rapid neutron capture (r-process nucleosynthesis) producing heavy elements. The radioactive decay of these heavy and unstable elements power the isotropic and thermal emission known as kilonova \citep{2013ApJ...776L..40Y, 2014MNRAS.441.3444M, 2017LRR....20....3M}. The kilonova ejecta is comparatively much slower than the jet and this mildly relativistic ejecta interacts with the ambient medium on the timescale of nearly a few years since the explosion depending on the energy injection from the central engine as well as the density of the medium \citep{2011Natur.478...82N}. If the resultant product of the merger is a millisecond magnetar, the ejecta would be energized because of the continuous energy injection from the magnetar through the spin down process. This energized ejecta inevitably crashes into the ambient medium resulting in a long lived synchrotron emission which peaks in the radio (cm) bands. Late time observations are therefore important to detect this emission which will not only help to get a complete picture of the central engine but will also allow to put constraints on the nature of the ambient medium \citep{2011Natur.478...82N, 2014MNRAS.437.1821M}. 

In the past, a few studies have been carried out by \citet{2014MNRAS.437.1821M}, \citet{2016ApJ...819L..22H}, \citet{2016ApJ...831..141F}, \citet{2020ApJ...902...82S} and \citet{Ricci2021} using the late time observations acquired with the Very Large Array (VLA) and Australian Telescope Compact Array (ATCA). All these observations were performed at a frequency of 1.4 GHz or more. The upper limits of flux density quoted in the study by \citet{2014MNRAS.437.1821M} were quite high and they concluded that the number density value of $n_0 \leq 10^{-1} cm^{-3}$ can be constrained for a stable magnetar remnant with the rotational energy of $E_{rot} \sim 10^{53}$ erg. The study by \citet{2016ApJ...819L..22H} and \citet{2016ApJ...831..141F} constrained the number density to be $n_0 \leq 10^{-3} cm^{-3}$ for a large magnetar rotational energy of $E_{rot} \sim 10^{53}$ erg and ejecta mass of $M_{ej} \sim 10^{-2} M_{\odot}$. In \citet{2020ApJ...902...82S}
and \citet{Ricci2021}, the ambient medium density was constrained between 2$\times 10^{-3}$ and 2$\times 10^{-1} cm^{-3}$ for the maximum rotational energy of magnetar $\leq 10^{52}$ erg and ejecta mass of $\leq 0.12 M_{\odot}$

We have carried out a comprehensive study with a sample of short GRBs upto the year 2017 which lie at a redshift of $z \leq 0.16$. Owing to the small distance where these bursts lie, they are ideal candidates for detecting any late time emission from the merger ejecta. We assume these events are the results of BNS merger and with the availability of detection upper limits our model will allow to put stringent constraints on the remnant and ambient medium properties. In our model, we use the basic magnetar model taking into account the energy injection from the central engine, evolution of the synchrotron frequencies, and generic hydrodynamics as well as propose modifications by incorporating the relativistic correction due to the evolution of the bulk Lorentz factor. A detailed discussion of our model is presented in section \ref{model}

The paper is organised as follows. In section 2 we discuss the sample selection of short GRBs. Section 3 describes our model and the modifications we have incorporated as compared to the other previous studies. In section 4 we provide model parameters for the selected bursts and conclude our findings in section 5. Throughout the paper, we assume a $\Lambda$CDM cosmology with $H_0= 70 \, \rm km \, s^{-1}\, Mpc^{-1}$, $\Omega_m = 0.27$ and $\Omega_{\Lambda} = 0.73$ \citep{Komatsu2011}.

\section{Sample Selection}
\label{sample}

We choose short GRBs in the local universe with spectroscopic redshift upto 0.16 and those showing a plateau phase in their X-ray lightcurves. The previous studies by \citet{2016ApJ...831..141F,2020ApJ...902...82S} and \citet{Ricci2021} present late time data of short GRBs which exibhit a plateau phase in their X-ray lightcurves. We select GRBs from these studies which have observations at 2.1 and 6.0 GHz. 17 GRBs in the literature show a plateau in their X-ray lightcurves. Among these only 7 GRBs lie within the redshift upto 0.16. GRB 170817A located at a very low redshift of 0.0097 is also included in our sample. The observational details of 7 GRBs in our sample are given in Table \ref{tab:radio_observation}.

\begin{table*}
\caption{Observational details of 7 GRBs in our sample}
\centering
\smallskip
\begin{tabular}{l c c c c c l}
\hline \hline
GRB  &Redshift       &Frequency         	&$T_{0bs}^{1}$    	    & 3$\sigma$ upper limit & X-ray behavior    &  Reference                    \\
         & (z)      &Band (GHz)	&  (years)  	    & ($\mu$Jy) &     &   \\
\hline         
060614  & 0.125 & 2.1  & 9.41 & 252 & Extended Emission &  \citet{2015NatCo...6.7323Y}  \\ 
            &       & 2.1  & 0.92 & 150 &   &  \citet{2016ApJ...819L..22H}  \\
061201  & 0.111 & 2.1 & 8.91 & 351 & &   \citet{stratta} \\
080905 &    0.122   & 2.1 & 7.37 & 537& Early Steep Decline &   \citet{2010MNRAS.408..383R}	\\
       &            & 6.0 & 5.77 & 61.8 & Early Steep Decline& \citet{2016ApJ...831..141F}  	\\
       &            & 6.0 & 6.47 & 57& &  \citet{2010MNRAS.408..383R}     \\
130822A	& 0.154& 2.1 & 2.18 & 270 & Early Steep Decline & \citet{Ricci2021}  	\\
        &      & 6.0 & 5.38 & 27& Early Steep Decline&  \citet{Ricci2021} \\
150101B	& 0.1341& 2.1 & 0.82 & 234& off-axis &  \citet{2018NatCo...9.4089T}  	\\
        &        & 6.0 & 3.03 & 99 & off-axis& \citet{2018NatCo...9.4089T}   \\
160821B  & 0.1613 & 6.0   & 2.38  & 18 & Early Steep Decline& \citet{2019MNRAS.489.2104T}  \\ 	
170817A  & 0.0097 & 6.0   & 1.98  & 24.2 & Extended Emission & \citet{2019ApJ...886L..17H}                 \\    
\hline
\multicolumn{6}{l}{$T_{obs}^{1}$ is calculated since burst trigger time.}\\
\end{tabular}
\label{tab:radio_observation}      
\end{table*}

\section{Magnetar Modeling}
\label{model}

The late time radio emission in short GRBs powered by magnetars is a result of the interaction between the tidally disrupted ejecta and the ambient medium \citep{2011Natur.478...82N}. The first consideration is that the ejecta is emerging spherically from the central engine with an initial velocity $\beta_0$. The emission from the ejecta is characterised by several parameters like ejecta mass $M_{ej}$, rotational energy of the magnetar $E_{rot}$, ambient medium density $n_0$, the fraction of post-shock energy into the accelerating electrons and magnetic field $\epsilon_e$ and $\epsilon_B$ respectively and the power-law distribution index of electrons `p' with $N(\gamma) \propto \gamma^{-p}$. The simplistic magnetar modeling has earlier been used in \citet{2011Natur.478...82N, 2014MNRAS.437.1821M,2016ApJ...819L..22H} and \citet{2016ApJ...831..141F} where they included the energy injection from the central engine and the synchrotron self absorption effect. Some modifications like generic hydrodynamics and deep Newtonian phase were proposed in \citet{2020ApJ...890..102L, 2020ApJ...902...82S} and \citet{Ricci2021}. However, none of these models consider the evolution of the bulk Lorentz factor. We discuss the magetar model in detail below incorporating the relativistic correction.

\subsection{Ejecta dynamics} 
The rotational energy of the stable NS remnant formed in a BNS merger can be calculated using

\begin{equation}
    E_{rot}=\frac{1}{2}I\Omega^2\simeq3\times10^{52} erg~\bigg(\frac{P}{1 ms}\bigg)^{-2}
\end{equation}

where I $\simeq$ 1.5 $\times$ $10^{15}$ g $cm^{-3}$ is the moment of inertia of a NS. In presence of a magnetar central engine, a significant fraction of its rotational energy is transferred to the ejecta which acts as a driving force behind the acceleration of the ejecta \citep{2007PhR...442..166N}. 

In order to calculate the deceleration radius and timescale, it is mandatory to consider the evolution of the bulk Lorentz factor ($\Gamma$) of the ejecta with the swept up mass ($M_s$) as described in \citet{2012ApJ...752L...8P} and is shown below: 

\begin{equation}
    \frac{d\Gamma}{dM_{s}} = -\frac{\hat{\gamma}(\Gamma^2 - 1) -(\hat{\gamma} - 1) \Gamma \beta^2}{M_{ej} + M_s[2 \hat{\gamma}\Gamma - (\hat{\gamma} - 1) (1 + \Gamma^{-2})]}
\end{equation}
where $\hat{\gamma}$ is the adiabatic index, $M_{ej}$ is the ejecta mass, and $\beta$ is the velocity of the ejecta.

The evolution of swept-up mass from ambient medium ($M_s$) and ejecta radius (r) are given in \citet{2012ApJ...752L...8P} as follows:

\begin{equation}
    \frac{dM_s}{dr} = 4\pi r^2n_0m_p
\end{equation}
and
\begin{equation}
    \frac{dr}{dt} = \Gamma^2 \beta(\Gamma) c (1 + \beta(\Gamma))
\end{equation}
where $m_p$ is the mass of proton and $n_0$ is the density of the ambient medium.

On solving equations 2, 3 and 4 we obtain expressions for $\Gamma$, swept-up mass $M_s$ and radius.

Initially, there is an enhancement in the kinetic energy of the ejecta because of continuous energy injection from the magnetar. When the ejecta collects a mass comparable to its own from the ambient medium, the ejecta starts decelerating at the characteristic timescale ($t_{dec}$) and radius ($R_{dec}$). In order to calculate $t_{dec}$ and $R_{dec}$, we have considered a spherical outflow from the central engine with the rotational energy ($E_{rot}$) of the magnetar and initial Lorentz factor ($\gamma_0$) with the associated velocity $c\beta_0$ which propagates through the CSM of density $n_0$.

\begin{equation}
R_{dec}\simeq \bigg(\frac{3E_{rot}}{4\pi n_0m_pc^2\gamma_0(\gamma_0 - 1)}\bigg)^{1/3}
\end{equation}

and

\begin{equation}
    t_{dec}\sim  \frac{R_{dec}(1 - \beta_0)}{c\beta_0}
\end{equation}

The minimum Lorentz factor ($\gamma_m$) and the magnetic field strength (B) in the shock can be calculated using the fraction of energy transferred to the electron energy and magnetic field ($\epsilon_e$ and $\epsilon_B$) respectively.

\begin{gather}
\label{elec}
    \gamma_{m} = 1 + \bigg(\frac{p-2}{p-1}\bigg) \frac{m_p}{m_e}\epsilon_e(\Gamma - 1)
\end{gather}

and

\begin{equation}
\label{mag}
     B = c (32 \pi n_0 m_p \epsilon_B \Gamma(\Gamma - 1))^{1/2}
 \end{equation}

\subsection{Calculation of synchrotron frequencies}
The late time radio spectrum is completely dominated by two frequencies. One of them is the typical synchrotron frequency  $\nu_m$ of the electrons having the minimum Lorentz factor $\gamma_m$. 

\begin{equation}
    \nu_m = \bigg(\frac{3q}{4\pi m_e c} \bigg) B \gamma_m^2 \Gamma (1+z)^{0.5}
\end{equation}
where {\it q} and $m_e$ are charge and mass respectively of the electron, {\it c} is the velocity of light, {\it B} is the magnetic field, and {\it z} is the redshift of the GRB.

The other is the synchrotron self-absorption frequency $\nu_a$. Radio emission becomes observable when the observed frequency is above $\nu_a$. $\nu_a$ plays a very significant role in low frequency radio regime. The expression for $\nu_a$ from  \citet{2011Natur.478...82N} is shown below:

\begin{equation}
    \nu_a \approx 1 GHz \bigg(\frac{R}{10^{17}}\bigg)^{\frac{2}{(p+4)}} \bigg(\frac{\epsilon_B}{0.1}\bigg)^{\frac{2+p}{2(p+4)}} \bigg(\frac{\epsilon_e}{0.1}\bigg)^{\frac{2(p-1)}{p+4}} n^{\frac{6+p}{2(p+4)}} \beta_0^{\frac{5p-2}{p+4}}
\end{equation}

Another important frequency in the synchrotron spectrum is the cooling frequency $\nu_c$. For the late time emission scenario, $\nu_c$ lies much above the observable frequency and $\nu_m$ and $\nu_a$.  The contribution of $\nu_c$ in the late time radio lightcurves is therefore not significant.

\subsection{Synchrotron flux calculation}

If the observing frequency is greater than $\nu_m$ and $\nu_a$, then the corresponding observed peak flux density can be calculated from \citet{1999ApJ...523..177W}. The specific flux $f_{\nu}$ of a single electron can be written as

 \begin{equation}
     F_{{\nu},e}= \frac{P_{{\nu},e}}{4\pi d_L^2}
 \end{equation}

where $P_{\nu,e}$ is the power emitted by a single electron per unit frequency in rest frame and $d_L$ is the distance to the source. $P_{{\nu}_m,e}$ can be written as (Wijers \& Galama 1999).

\begin{equation}
     P_{{{\nu}_m,e}}= \phi_p\frac{\sqrt{3}e^3 B}{m_e c^2}
 \end{equation}
 
where $\phi_p$ is the flux of the dimensionless maximum point of the spectrum ($f(x_p) = \phi_p$). e and $m_e$ are representing the electron charge and mass respectively.


 
After accounting for the redshift and relativistic transformation, the total specific flux, $F_{{{\nu}_m},obs}$, can be written as

\begin{equation}
     F_{{{\nu}_m,obs}} = \frac{N_e P_{{\nu_m,e}} \Gamma (1+z)}{4 \pi d_L^2}
 \end{equation}
 
Here we are considering that the ejecta is ploughing through the constant density inter-stellar medium (ISM), then we can write $N_e = \frac{4}{3} \pi R^3 n$. Substituting $N_e$ and equation 12 in equation 13, we evaluate the total specific flux as

\begin{equation}
     F_{{{\nu}_m,obs}} = \frac{\sqrt{3} B e^3}{m_e c^2}\frac{R^3 n_0 \Gamma (1+z)}{3 d_L^2}
 \end{equation}
 
Substituting equation 8 in equation 14, we obtain the final flux equation in terms of {\it n} and $\epsilon_B$.

\begin{equation}
     F_{{{\nu}_m,obs}} = \frac{\sqrt{96 \pi n_0^3m_p\epsilon_B\Gamma^3(\Gamma - 1)} }{m_e c}\frac{R^3 e^3 (1+z)}{3 d_L^2}
 \end{equation}
 
\section{Lightcurve Analysis}
\label{analysis}

\begin{figure*}[h!]
\includegraphics[width=\textwidth]{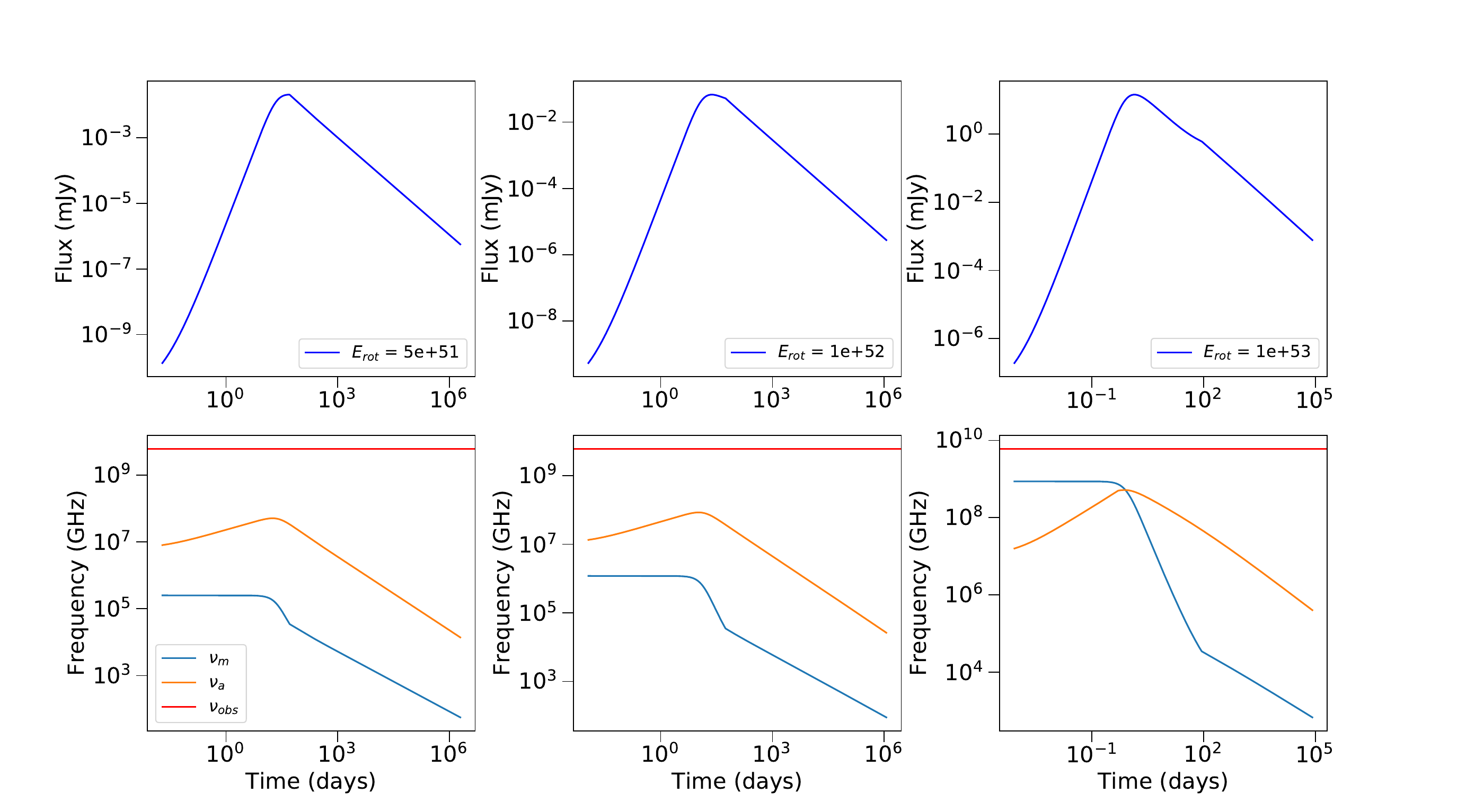}
\caption{Model lightcurves and the evolution of $\nu_a$ and $\nu_m$. The ejecta mass is fixed at 0.04 $M_{\odot}$ and number density of the ambient medium is fixed at $10^{-2} cm^{-3}$. The solid horizontal lines in the bottom panels indicate the observed frequency.}
\label{freq_evolution}
\end{figure*}

Using the above formulation we generate the model lightcurves and compare them with the 3$\sigma$ upper limits. While constructing the model lightcurves we take into account the assumptions and  and fiducial values of certain parameters described below.

\begin{itemize}
    \item Numerical simulations showed that long-lived merger remnants eject a large fraction of the remnant accretion disk mass \citep{2014MNRAS.441.3444M}. We consider ejecta mass values of 0.04 $M_{\odot}$ and 0.1 $M_{\odot}$ which corresponds to velocities $\beta \sim$ 0.5 and 0.3 respectively. Velocities below $\beta <$ 0.3 are discarded because they will produce very weak and delayed radio emission.
    \item We varied the magnetar rotational energy from $10^{51}$ to $10^{53}$ erg. The upper limit of rotational energy corresponds to a stable neutron star of mass 2.2 $M_{\odot}$ \citep{2015MNRAS.454.3311M}.
    \item The fraction of post shock energy into the accelerating electron $\epsilon_e$ is fixed at 0.1
    \item The fraction of post shock energy into the magnetic field $\epsilon_B$ is at 0.1 and 0.01. 
    \item The powerlaw index p is fixed at 2.4.
    \item The density of the ambient medium is not fixed. It generally varies in a vast range for different GRBs. We take the number density range between $10^{-5}$ to $1$ $cm^{-3}$ in six equal intervals.
\end{itemize}

\begin{figure*}[h!]
\includegraphics[width=\textwidth]{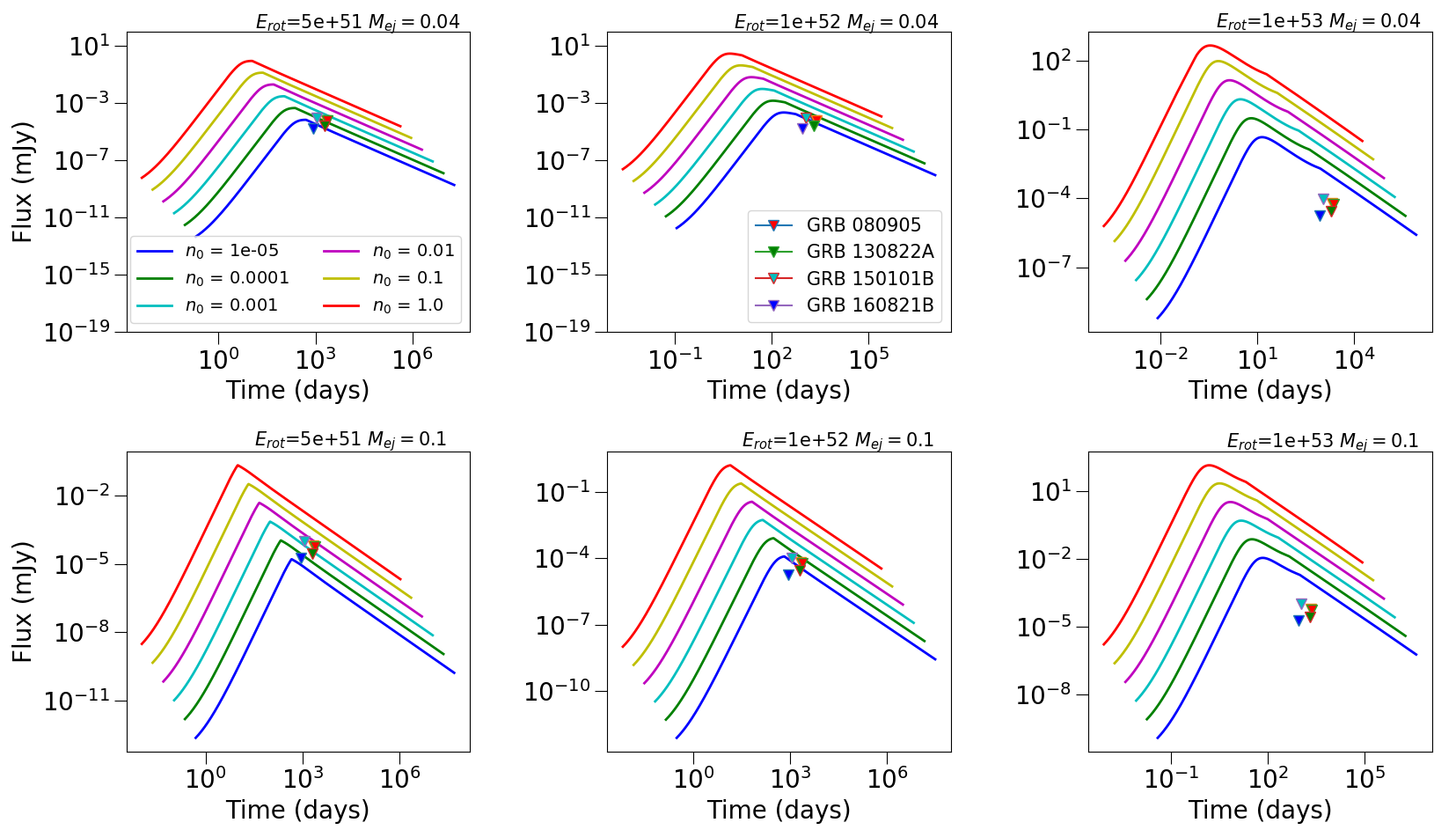}
\caption{Model lightcurves along with the VLA 6 GHz 3$\sigma$ upper limits. The upper limits are shown as inverted triangles in the figure. The value of $\epsilon_B$ is fixed to 0.1. For two different values of ejecta mass 0.04 and 0.1 $M_{\odot}$ we generate the model light curves with different magenetar rotational energy and number density as indicated in the plot.}
\label{fig:jaa_composite1}
\end{figure*}

\begin{figure*}[h!]
\includegraphics[width=\textwidth]{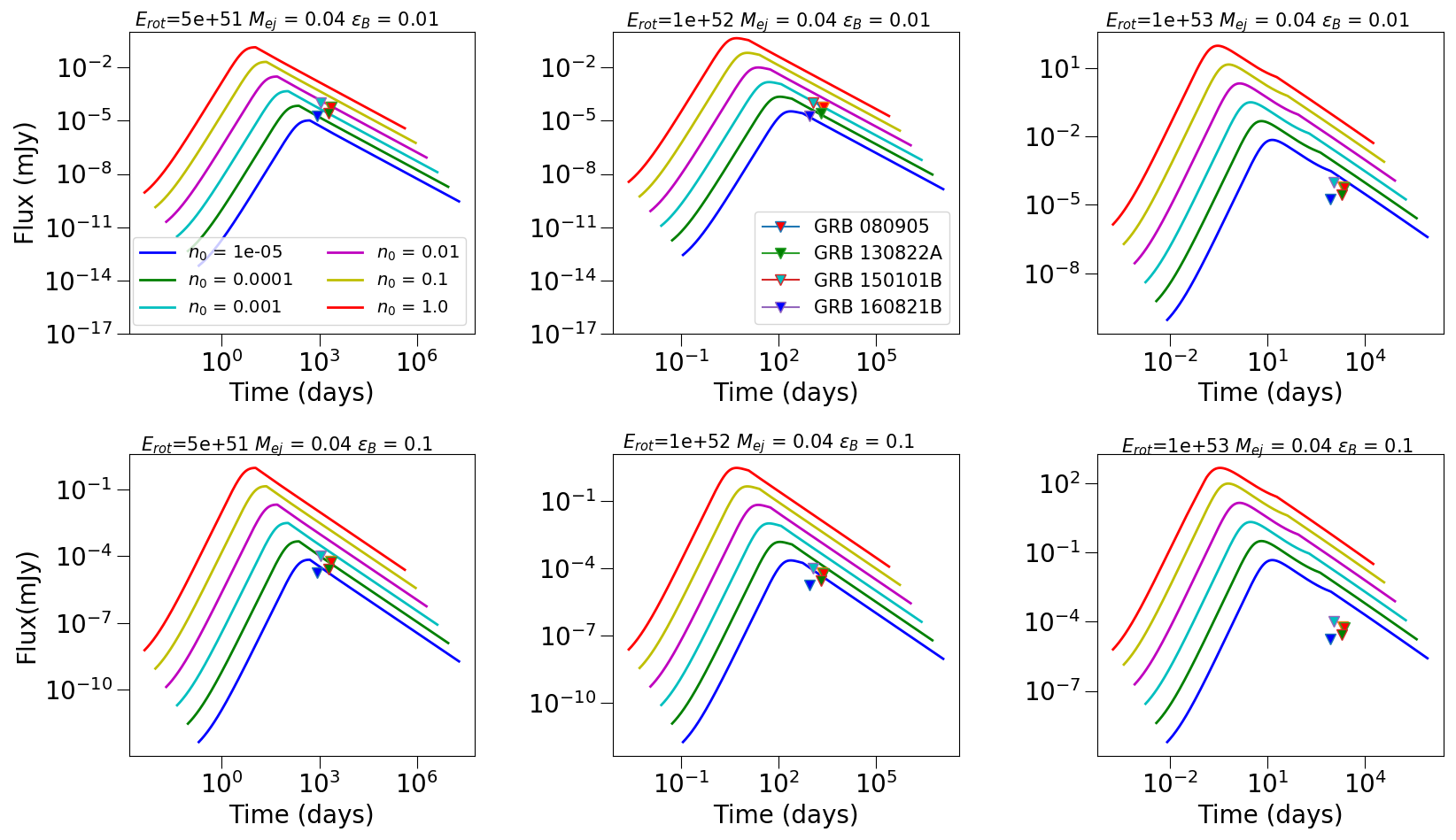}
\caption{Same as figure \ref{fig:jaa_composite1} but for a fixed value of ejecta mass and two different values of $\epsilon_B$ (0.01 and 0.1).}
\label{fig:jaa_composite2}
\end{figure*}

The top panel of figure \ref{freq_evolution} shows the model lightcurves for different values of rotational energy (5$\times 10^{51}$ -- $10^{53}$ erg) for a fixed ejecta mass (0.04 $M_{\odot}$) and number density ($10^{-2} cm^{-3}$) at a frequency of 6GHz. The evolution of $\nu_a$ and $\nu_m$ is shown in the bottom panel. $\nu_m$, which represents the average kinetic energy of the radiating electrons remains below $\nu_a$ for lower values of rotational energy. But for higher rotational energy, the electron population achieves higher $\gamma_m$, leading to $\nu_m$ going above $\nu_a$. The peak of the model lightcurves is consistent with the peak of $\nu_m$. At late times the role of $\nu_c$ is insignificant as it lies much beyond the observable frequencies.

\begin{figure*}
\includegraphics[width=\textwidth]{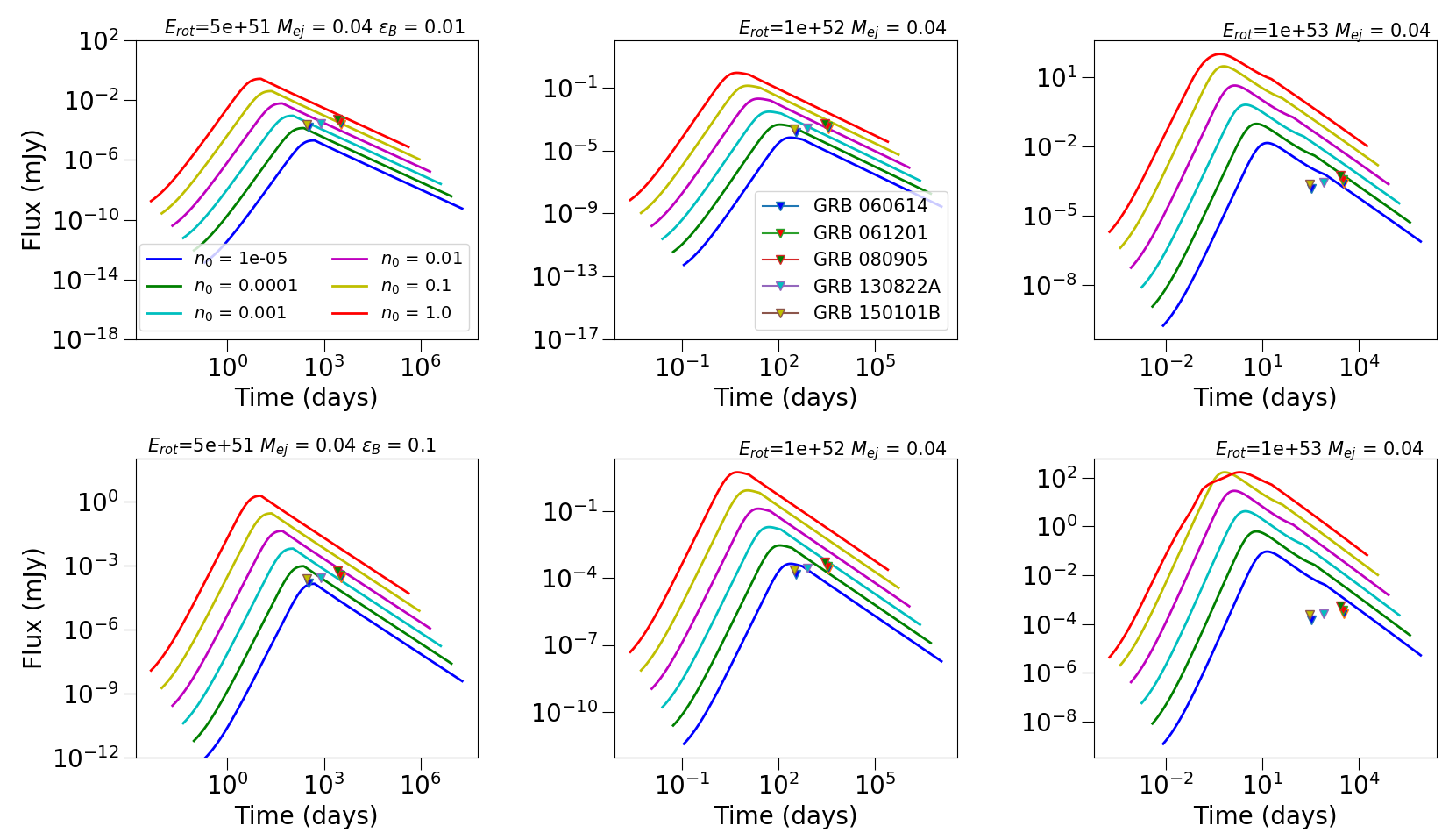}
\caption{Model lightcurves along with the ATCA 2.1 GHz 3$\sigma$ upper limits. The model lightcurves are generated for a fixed ejecta mass and two different values of $\epsilon_B$ (0.01 and 0.1).}
\label{ATCA}
\end{figure*}

Figures \ref{fig:jaa_composite1}, \ref{fig:jaa_composite2} and \ref{ATCA} shows the model lightcurves along with the 3$\sigma$ upper limits from VLA and ATCA of the GRBs in our sample (Table \ref{tab:radio_observation}). The models are generated for different values of ejecta mass, $\epsilon_B$, magnetar rotational energy and number density as indicated in the figures. The model lightcurves in these figures depict that the peak time of the lightcurves are decreasing with increasing values of rotational energy of the magnetar and the ambient medium density. In addition to that, the model flux decreases for increasing values of ejecta mass and decreasing values of magnetar rotational energy. 

From figures \ref{fig:jaa_composite1} and \ref{fig:jaa_composite2} it is evident that the most energetic scenario ($E_{rot} \sim 10^{53}$ erg) is disfavoured as the upper limits lie beyond the model lightcurves and would require much smaller ambient medium density ($< 10^{-5} cm ^{-3}$) and $\epsilon_B$ ($< 0.01$) to satisfy the observational constraints. In a study by \citet{Ricci2021} it was found that the number densities are better constrained for GRBs at lower redshifts as compared to those lying at higher redshifts. In our sample of low redshift short GRBs, the models with ambient medium densities of $n_0 \geq 10^{-4} cm^{-3}$ are ruled out for magnetar rotational energy of $10^{52}$ erg. Models with  rotational energy $\sim 5 \times 10^{51}$ erg constrains the number density values within $10^{-5}$ to $10^{-3} cm^{-3}$. The number density obtained in this work are consistent with the previous studies by \citet{Ricci2021}.

The model lightcurves with high rotational energy ($10^{53}$ erg) and lower value of $\epsilon_B$ = 0.01 are satisfied only for very low number density values ($n_0 \leq 10^{-5} cm^{-3}$). However, for lower magnetar rotational energy ($10^{51}$ and $10^{52}$ erg), stringent constraints for number density values (between $10^{-4}$ to $10^{-2} cm^{-3}$) have been found.

\begin{figure*}
\includegraphics[width=\textwidth]{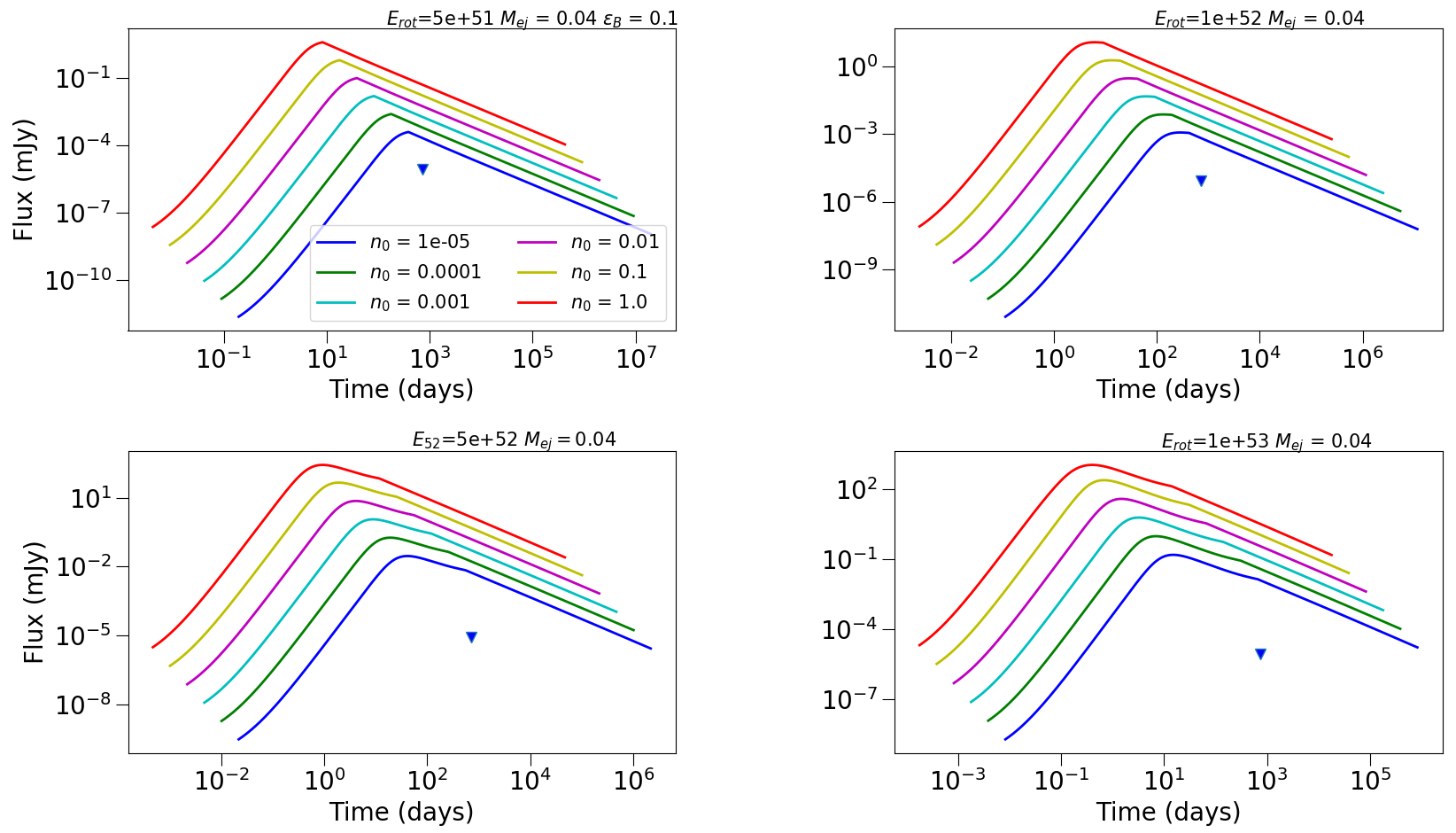}
\caption{Model lightcurves along with the VLA 6 GHz 3$\sigma$ upper limit of GRB~170817A. The ejecta mass and $\epsilon_B$ values are fixed at 0.04 $M_{\odot}$ and 0.1 respectively.}
\label{GRB170817A}
\end{figure*}

In figure \ref{GRB170817A} we show the model lightcurves, generated for a fixed value of ejecta mass and $\epsilon_B$ considering different values of rotational energy and number density, along with the 3$\sigma$ upper limit of GRB~170817A. The upper limit available is at $\sim$714 days since burst. The available upper limit does not satisfy any of the models indicating that different values of the parameters may be required to generate the models. Further late time monitoring of GRB~170817A would be useful to search for the emission at late times due to merger ejecta.

\section{Conclusions}
A faint radio emission is expected when the merger ejecta collides with the ambient medium in presence of a magnetar central engine \citep{2011Natur.478...82N}. Several studies in the past have reported upper limits in a quest to search for merger ejecta emission in radio bands from short GRBs \citep{2014MNRAS.437.1821M, 2016ApJ...819L..22H, 2020ApJ...902...82S, 2016ApJ...831..141F, Ricci2021}. The upper limits are very useful to constrain the ambient medium density and the rotational energy of the magnetar. We select a sample of 7 short GRBs located at a redshift of $z \leq 0.16$ for which upper limits were reported in the literature.  We used the standard magnetar model and modified it to incorporate the relativistic corrections. The model light curves thus generated are used to constrain the basic parameters of the central engine and ambient medium.

We find that models with high rotational energy are disfavoured whereas models with lower values of rotational energy provide tight constraints on the ambient medium density ($n_0 \sim 10^{-5} - 10^{-3} cm^{-3}$) except the ATCA 2.1 GHz lighcurves for $\epsilon_B = 0.01$. The number density values obtained are very low.  It is also seen that changing the ejecta mass does not have much effect on the late time lightcurves. The radio emission from merger ejecta is expected to peak at the deceleration time (typically after a few years since the burst). Therefore, observations not very far out in time will not be able to detect any radio emission, if any. It is important to acquire observations at late time. In the case of GRB~170817A the model lightcurves are unable to explain the radio upper limits. Late time observations of GRB~170817A could substantially refine these constraints. 


In the near future, it may be possible to detect merger ejecta emission from near by local short GRBs with the next generation radio telescopes and prove the existence of magnetar central engine. The upcoming radio telescopes like the Square Kilometer Array (SKA) and New Generation VLA (ng-VLA) with increased sensitivity of $\mu$Jy level will push the detection limits of merger ejecta emission at late time.

\vspace{2em}







\section*{Acknowledgements}

KM acknowledges BRICS grant {DST/IMRCD/BRICS/PilotCall1/ProFCheap/2017(G)} for the financial support. KGA is partially supported by the Swarnajayanti Fellowship Grant No.DST/SJF/PSA-01/2017-18, MATRICS grant MTR/2020/000177 of SERB, and a grant from the Infosys Foundation.

\vspace{-1em}


\bibliography{jaa} 


\end{document}